\documentclass[conference]{IEEEtran}
\IEEEoverridecommandlockouts
\usepackage{xcolor} 
\usepackage{cite} 
\usepackage{amsmath,amssymb,amsfonts,bbm} 
\usepackage{algorithmic} 
\usepackage{tabularx} 
\usepackage{amsmath, bm} 
\usepackage{amsmath,autobreak} 
\usepackage[ruled, linesnumbered]{algorithm2e} 
\usepackage{subfigure} 
\usepackage{soul} 
\usepackage{balance} 
\usepackage{color} 
\usepackage{float} 
\usepackage{autobreak} 
\usepackage{graphicx} 
\usepackage{textcomp} 
\usepackage{xcolor} 
\usepackage{booktabs} 
\usepackage{multirow} 
\usepackage{bbm} 
\usepackage{dsfont} 
\usepackage{footnote} 
\usepackage[]{mathtools} 
\usepackage{lipsum} 
\makesavenoteenv{algorithm} 
\usepackage{diagbox}
\usepackage{lineno}
\usepackage{tikz}
\usepackage[colorlinks, linkcolor=blue, anchorcolor=blue, citecolor=blue]{hyperref}

\newcommand*{\circled}[1]{\lower.7ex\hbox{\tikz\draw (0pt, 0pt)%
    circle (.5em) node {\makebox[1em][c]{\small #1}};}}

\def\BibTeX{{\rm B\kern-.05em{\sc i\kern-.025em b}\kern-.08em
    T\kern-.1667em\lower.7ex\hbox{E}\kern-.125emX}}
\begin{document}

\title{Graph Neural Network-Based Bandwidth Allocation for Secure Wireless Communications}

\author{\IEEEauthorblockN{
Xin Hao, 
Phee Lep Yeoh,
Yuhong Liu,
Changyang She,
Branka Vucetic, 
and Yonghui Li
}

\IEEEauthorblockA{School of Electrical and Information Engineering, The University of Sydney, Australia}
}

\maketitle

\begin{abstract}
This paper designs a graph neural network (GNN) to improve bandwidth allocations for multiple legitimate wireless users transmitting to a base station in the presence of an eavesdropper. To improve the privacy and prevent eavesdropping attacks, we propose a user scheduling algorithm to schedule users satisfying an instantaneous minimum secrecy rate constraint. Based on this, we optimize the bandwidth allocations with three algorithms namely iterative search (IvS), GNN-based supervised learning (GNN-SL), and GNN-based unsupervised learning (GNN-USL). We present a computational complexity analysis which shows that GNN-SL and GNN-USL can be more efficient compared to IvS which is limited by the bandwidth block size.  
Numerical simulation results highlight that our proposed GNN-based resource allocations can achieve a comparable sum secrecy rate compared to IvS with significantly lower computational complexity. Furthermore, we observe that the GNN approach is more robust to uncertainties in the eavesdropper's channel state information, especially compared with the best channel allocation scheme. 

\end{abstract}

\section{Introduction}
Machine learning has recently been demonstrated to be a powerful approach for optimizing resource allocations in 5G-and-beyond  wireless networks~\cite{SCY_tutorial_urllc}. Compared to conventional numerical optimizations requiring lengthy iterations to find the optimal policy, machine learning algorithms can be trained offline to approximate the optimal policy, thus making it more efficient for real-time resource allocation for users with frequently changing wireless channels~\cite{SCY_digital_twin}. 
A significant amount of computational cost for machine learning involves collecting training labels using the typical supervised learning (SL) approach~\cite{outperforms_random_SVM_etc}. 
To further reduce the computational cost, unsupervised learning (USL) approaches without training labels have been developed for optimizing resource allocations in wireless networks~\cite{LYH_GNN_interference}. 

Recent works have considered the use of machine learning to optimize resources in industrial wireless networks with strict physical-layer security requirements~\cite{industry_Secure_ML}. The eavesdropping attack is a well-known physical-layer security attack, which jeopardizes the user's privacy by wiretapping the transmitted information~\cite{Ring_shape_location_Eve}, and is evaluated based on the secrecy rate according to the channel conditions of the legitimate user and eavesdropper~\cite{TWC_SR_Poor}. In~\cite{thz_SR_GC}, an SL-based deep neural network (DNN) was used to maximize the secrecy rate of a legitimate user and eavesdropper by jointly optimizing the power allocation, carrier frequency, transmit power, and waveform. In~\cite{tvt_RS_non_robust}, the authors applied a DNN to optimize the transmit power allocation aimed at maximizing the secrecy rate of a single user subject to an interference leakage threshold constraint. In~\cite{INFOCOM_SR_DRL}, the authors used deep reinforcement learning to optimize the downlink transmit power aimed at maximizing the average secrecy rate of multiple legitimate users with a single eavesdropper.
We note that the DNN-based approach is effective for finding optimal solutions in wireless networks with dynamic channels but is limited to a fixed number of users with a {statistic}  constraint for {the} users~\cite{TWC_max_sum_rate}. Recent works have considered using graph neural networks (GNNs) to support both dynamic wireless channels and a dynamic number of users with {instantaneous} constraints~\cite{MPNN_chemistry}. 
In~\cite{MPGNN_HK_JSAC}, the scalability of the neuron numbers was successfully achieved by updating the GNN's parameters using the aggregated feature.

In this paper, we apply GNNs to optimize the bandwidth allocations for multiple legitimate users transmitting to a base station (BS) in the presence of an eavesdropper with dynamic locations. Our proposed  GNN-based optimization framework eliminates eavesdropping attacks by satisfying an instantaneous constraint on the secrecy rate for each user, thus introducing an extra dynamic requirement on the number of users.
The main contributions are summarized as follows:
\begin{itemize}
\item We protect user's communication privacy by preventing potential eavesdropping attacks based on an instantaneous constraint on the minimum secrecy rate of each user in our proposed user scheduling algorithm. 
We consider the practical scenario of dynamic eavesdropper locations in the optimization formulation.
\item {The iterative search algorithm is applied to find the optimal bandwidth allocation strategy for each possible channel condition. Then, we} design a GNN with a dynamic number of neurons to represent the dynamic bandwidth allocation problem. {We propose both GNN-based SL and USL algorithms which can achieve comparable performance with IvS.} 
\item The computational complexities of our proposed GNN-USL and benchmark bandwidth allocation algorithms are analyzed in detail. {The GNN-USL algorithm reduces computations for collecting labels compared with GNN-SL.} The robustness of our proposed algorithm is also highlighted in the simulations by considering the practical scenario of the eavesdropper's CSI uncertainty. 
\end{itemize}
Numerical and simulation results show that our proposed {GNN-SL} and GNN-USL solutions achieve comparable sum secrecy rates to the iterative search approach with significantly lower computational complexity. Furthermore, compared with iterative search and best channel bandwidth allocation algorithms, our GNN-SL and GNN-USL approaches can guarantee the secrecy rate with low computational complexity. 


\begin{figure}[t]
\centering
{\includegraphics[scale=0.8]{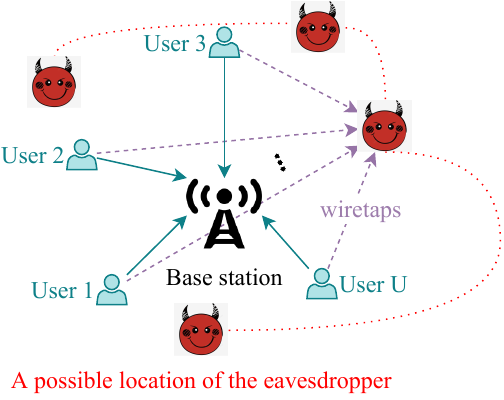}} 
\caption{$U$ legitimate users transmit confidential information to a BS, and a strong eavesdropper intends to wiretap the transmitted information.}
\label{fig_systemModel_UL}
\end{figure}

\section{System Model}
Fig.~\ref{fig_systemModel_UL} shows our considered system model with $U$ randomly located legitimate users transmitting confidential information to the BS in the presence of an eavesdropper. We aim to maximize the sum secrecy rates of all the users by optimizing the bandwidth allocated to individual users. We assume the eavesdropper is constantly changing its location and is able to wiretap all the bandwidths of legitimate users. 

\subsection{Secrecy Rate}
The data rate of the $u$-th user can be expressed as
\begin{equation}
R_{u}^\mathrm{B} ={W_u}\log_2\left(1+ \frac{P_u(d_{u}^\mathrm{B})^{-\alpha} g_{u}^\mathrm{B} } { N_0   W_u}\right), 
\label{eq_user_data_rate}
\end{equation}
where $W_u$ is the bandwidth allocated to the $u$-th user, $P_u$ is the transmit power of the $u$-th user, $d_{u}^\mathrm{B}$ is the distance from the $u$-th user to the BS, $\alpha$ is the path loss exponent, $g_{u}^\mathrm{B}$ is the small-scale channel gain of the legitimate channel for the $u$-th user, and $N_0$ is the single-sided noise spectral density. Similarly, the wiretapped data rate of the $u$-th user is given by
\begin{equation}
R_{u}^\mathrm{E}=  {W_u}\log_2\left(1+ \frac{P_u {(d_u^{\mathrm{E}})}^{-\alpha} g_u^{\mathrm{E}}}{ N_0  W_u}\right), 
\label{eq_eavesdropper_rate}
\end{equation}
where $d_u^\mathrm{E}$ is the distance from the $u$-th user to the eavesdropper, and $g_{u}^\mathrm{E}$ is the small-scale channel gain of the wiretapped channel for the $u$-th user. The secrecy rate of the $u$-th user can be described as
\begin{equation}
R_u^\mathrm{S} = \left[R_u^\mathrm{B} - R_{u}^\mathrm{E} \right]^{+}, 
\label{eq_secrecy_rate}
\end{equation}
where $[x]^{+}=\max\{0,x\}$. 

To eliminate eavesdropping attacks on any of the users, we require the secrecy rate of each user is above a minimum threshold, i.e.,
\begin{equation}
R_u^\mathrm{S} \geq R_{\min}^\mathrm{S},
\label{eq_Rs_min_constraint}
\end{equation}
where $R_{\min}^\mathrm{S} > 0$ is the required secrecy rate threshold.


\subsection{Optimization Problem Formulation}
We define the channel matrix of all the users as $\boldsymbol{H}^\mathrm{U} = [\boldsymbol{h}_1, \boldsymbol{h}_2, \cdots, \boldsymbol{h}_u ,\cdots, \boldsymbol{h}_U]^\mathrm{T}$, where $\boldsymbol{h}_u =[d_u^\mathrm{B}, d_u^\mathrm{E}, g_u^\mathrm{B}, g_u^\mathrm{E}]^\mathrm{T}$ is a vector representing the CSI of the $u$-th user. In our model, 
the allocated bandwidth for each user is a function of the corresponding CSIs. Thus, we have $\boldsymbol{W}^\mathrm{U}(\boldsymbol{H}^\mathrm{U})=[W_1(\boldsymbol{h}_1), \cdots, W_U(\boldsymbol{h}_U)]^\mathrm{T}$. Similarly, the secrecy rate of each user is a function of the allocated bandwidth, thus $R_u^{S}(W_u(\boldsymbol{h}_u)), \forall u\in \mathcal{U}$. We aim to maximize the sum secrecy rate of all the users as
\begin{align} 
\max_{\boldsymbol{W}^\mathrm{U}(\boldsymbol{H}^\mathrm{U})}   
&\sum\limits_{u =1}^{U}  
{R_u^\mathrm{S}  (W_u(\boldsymbol{h}_u))} ,   \label{eq_problem_function} \\
\mathrm{s.t.}\quad \nonumber
&\sum\nolimits_{u=1}^{U} {W_u(\boldsymbol{h}_u)} \leq W_{\mathrm{B},\max},      \tag{\ref{eq_problem_function}{a}} \label{eq_problem_func_a}\\
& W_u(\boldsymbol{h}_u)\geq 0,  \tag{\ref{eq_problem_function}{b}} \label{eq_problem_func_b}
\\
&R_u^\mathrm{S}(W_u(\boldsymbol{h}_u))\geq R_{\min}^\mathrm{S}, \tag{\ref{eq_problem_function}{c}} \label{eq_problem_func_c}
\end{align}
where $W_{\mathrm{B},\max}$ is the total available bandwidth at the BS.

\begin{algorithm}[t] 
\algsetup{linenosize=\normalsize} \small  
\caption{User Scheduling Algorithm.}\label{algorithm_user_scheduling}
Update the set of scheduled users: $\mathcal{K}=\mathcal{U}$.\\
$\circled{1}$\textit{Drop users that cannot satisfy the secrecy rate constraint with the maximum bandwidth}:\\
\For{$u \in \mathcal{U}$}
{
Calculate the secrecy rate of each user by taking $W_{\mathrm{B},\max}$ into~\eqref{eq_user_data_rate},~\eqref{eq_eavesdropper_rate}, and~\eqref{eq_secrecy_rate} as:
$R_{u,\max}^\mathrm{S} = [R_u^\mathrm{B} (W_{\mathrm{B},\max}) - R_{u}^\mathrm{E} (W_{\mathrm{B},\max}) ]^{+}$.\\
\uIf{$R_{u,\max}^\mathrm{S} = 0$}
{
Drop the $u$-th user: $\mathcal{K} = \mathcal{K} - \{u\}$, $K = K-1$, and $W_u = 0$.\\
}
\Else
{Get $W_{k,\min}$ using binary search:
$W_{k,\min} = f_\mathrm{BiS}(\boldsymbol{h}_k, R_{\min}^\mathrm{S})$. \label{algorithm_user_scheduling_line_W_k_min}
}
}
$\circled{2}$\textit{Drop users if there is insufficient bandwidth}:\\
\While{$\sum\nolimits_{k =1}^{K} {W_{k,\min}} > W_{\mathrm{B},\max}  $}
{
Identify user holding highest $W_{k,\min}$ in $\mathcal{K}$:
$k_{\mathrm{drop}} = \arg\max\limits_{k} {W_{k, \min}}$.\\
Drop the $k_\mathrm{drop}$-th user: $\mathcal{K} = \mathcal{K} - \{k_{\mathrm{drop}}\}$, $K = K-1$, and $W_{k_{\mathrm{drop}}}=0$.\\
}
\end{algorithm}

\section{Bandwidth Allocation Optimization}
In this section, we first introduce our user scheduling algorithm, which satisfies an instantaneous requirement on the secrecy rate. Next, we optimize bandwidth allocation using iterative search, which is identified as the optimal solution. Lastly, we design a GNN and utilize supervised and unsupervised learning algorithm to optimize the bandwidth allocation, respectively.

\subsection{User Scheduling}
\begin{figure}[t]
\centering
{\includegraphics[scale=0.8]{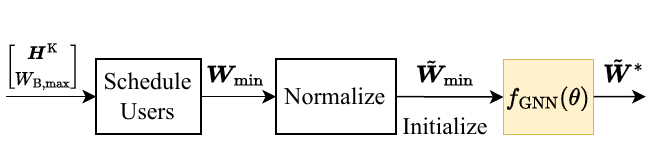}} 
\caption{The flow diagram of user scheduling and GNN-based bandwidth allocation.}
\label{fig_flow_schedule}
\end{figure}

Due to the dynamics of the wireless channels, the secrecy rate of each user are also dynamic. To resist the eavesdropping attacks in networks, we propose to schedule the users satisfying constraint~\eqref{eq_problem_func_c} for each possible channel condition. The specific user scheduling algorithm is detailed in Algorithm~\ref{algorithm_user_scheduling}. After user scheduling, problem~\eqref{eq_problem_function} is updated as follows
\begin{align} 
\max_{\boldsymbol{W}^\mathrm{K}(\boldsymbol{H}^\mathrm{K})} 
&\sum\limits_{k =1}^{K}  R_k^\mathrm{S} \left(W_k(\boldsymbol{h}_k)\right), \label{eq_positive_Rs_basic} 
\\
\mathrm{s.t.}\quad \nonumber
&\sum\nolimits_{k=1}^{K} {W_k(\boldsymbol{h}_k)} \leq W_{\mathrm{B},\max},      \tag{\ref{eq_positive_Rs_basic}{a}} \label{eq_positive_Rs_basic_a}
\\
& W_k({h}_k)\geq W_{k,\min},  \forall k \in \mathcal{K},															\tag{\ref{eq_positive_Rs_basic}{b}} \label{eq_positive_Rs_basic_b}
\end{align}
%
%
where $K \leq U$ is the number of scheduled users, $\boldsymbol{H}^\mathrm{K} = [\boldsymbol{h}_1, \cdots, \boldsymbol{h}_K]^\mathrm{T}$ is the channel matrix of the scheduled users, and $\boldsymbol{W}^\mathrm{K}(\boldsymbol{H}^\mathrm{K}) = [W_1(\boldsymbol{h}_1), \cdots, W_K(\boldsymbol{h}_k) ]^\mathrm{T}$ indicates the bandwidth allocated to the scheduled users. It is challenging to achieve the optimal solution since the time-varying variable $K$ introduces extra dynamics to problem~\eqref{eq_positive_Rs_basic}. Thus, we consider a GNN-based algorithm to solve problem~\eqref{eq_positive_Rs_basic}. The flow diagram is depicted in Fig.~\ref{fig_flow_schedule}, where $\boldsymbol{W}_{\min} = [W_{1,\min}, \cdots, W_{K,\min}]^\mathrm{T}$ indicates the vector of the minimum bandwidth required to satisfy constraint~\eqref{eq_Rs_min_constraint}, and is achieved in line~\ref{algorithm_user_scheduling_line_W_k_min} of Algorithm~\ref{algorithm_user_scheduling}. $\boldsymbol{\tilde{W}}_{\min} = [\tilde{W}_{1,\min}, \cdots, \tilde{W}_{K,\min}]^\mathrm{T} =[W_{1,\min}, \cdots, W_{K,\min}]^\mathrm{T} /{W_{\mathrm{B},\max}} = {\boldsymbol{W}_{\min}}/{W_{\mathrm{B},\max}}$ is a vector represents the normalized value of $\boldsymbol{W}_{\min}$.

\subsection{Bandwidth Allocation Algorithms}

\subsubsection{Iterative Search}
Motivated by~\cite{SCY_transfer}, we apply an iterative search (IvS) algorithm in our model to find the optimal bandwidth allocation policy. We have proved in Appendix~\ref{appendix_concave} that problem~\eqref{eq_positive_Rs_basic} is concave, thus IvS can find the optimal bandwidth allocation policy\footnote{The authors in~\cite{SCY_transfer} had proved the optimality if the two conditions holds. Since a convex problem satisfies both the conditions, we do not prove repetitive proof of optimality in this paper.}. Specifically, we use a resource block of the bandwidth, $\Delta W$, to calculate the secrecy rate incremental of each user, $\Delta R_k^\mathrm{S}$. Then IvS allocates an extra amount of bandwidth of $\Delta W$ to the user holding the highest $\Delta R_k^\mathrm{S}$. The step-by-step algorithm of IvS is detailed in Algorithm~\ref{algorithm_bandwidth_allocation_incremental_search}. 
Different from~\cite{SCY_transfer} which considered discretized resource allocations, we consider continues bandwidth resources. Thus, a small value of the resource block, $\Delta W$, is essential to guarantee the optimality of the bandwidth allocation policy. 

\begin{algorithm}[t]
\algsetup{linenosize=\normalsize} \small  
\caption{Bandwidth Allocation by IvS.}\label{algorithm_bandwidth_allocation_incremental_search}
\textbf{Initialize}: Size of resource block: {$\Delta W$}. \\
$W_k = W_{k,\min}, \forall k \in \mathcal{K}$.\\
\While{$ W_{\mathrm{B}, \max} - \sum\nolimits_{k =1}^{K} {W_k} \geq \Delta W$}
{
\For{$k \in \mathcal{K}$}
{
$\Delta R_k^\mathrm{S} (W_k)= R_k^\mathrm{S}(W_k + \Delta W) - R_k^\mathrm{S} (W_k)$.
}
Identify user has highest $\Delta R_k^\mathrm{S}(W_k)$ in $\mathcal{K}$: $k_{\mathrm{allo}} = \arg\max\limits_{k} {\Delta R_k^\mathrm{S} }(W_k)$.\\
Allocate extra $\Delta W$ bandwidth to the $k_\mathrm{allo}$-th user: $W_{k_\mathrm{allo}} = W_{k_\mathrm{allo}} + \Delta W$.
}
\textbf{Output}: $\boldsymbol{W}^*= \boldsymbol{W}_{\mathrm{B},\max}$
\end{algorithm}

\subsubsection{GNN Design}
\begin{figure}[t]
\centering
{\includegraphics[scale=0.8]{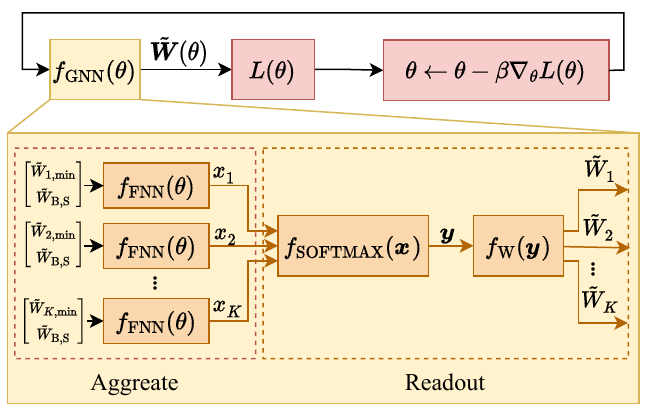}}
\caption{Designed GNN-based bandwidth allocation optimization algorithm.}
\label{fig_gnn_layer}
\end{figure}
%
%
%
We denote each scheduled user as a vertex in the GNN, and use a fully connected neural network (FNN) to extract the feature of each vertex. The flow diagram of the designed GNN-based bandwidth allocation algorithm is depicted in Fig.~\ref{fig_gnn_layer}. We denote the feature of the $k$-th vertex as $x_k$, and the FNN approximates the feature according to 
\begin{equation}
x_k = f_{\mathrm{FNN}}\left(\tilde{W}_{k,\min}, \tilde{W}_{\mathrm{B,S}}; \theta\right),
\label{eq_hidden_feature}
\end{equation}
where function $f_{\mathrm{FNN}}(\cdot)$ represents an FNN parameterized by $\theta$. Since the channels of different users follow the same distribution, we use the same parameter to get the feature, $x_k$, for different vertices. The $\tilde{W}_{\mathrm{B,S}} =1-\sum_{k=1}^{K} \tilde{W}_{k,\min}$ is the normalized surplus bandwidth. To optimize the bandwidth allocated to each scheduled user, we aggregate the features of scheduled users by using concatenation function,
\begin{equation}
\boldsymbol{x} =f_\mathrm{CONCAT}\left(x_1, \cdots, x_K\right)= \left[x_1, \cdots, x_K\right]^\mathrm{T}.
\end{equation}
The normalized vector is achieved by a \textit{softmax} activation function as
\begin{equation}
\boldsymbol{y} = f_\mathrm{SOFTMAX}\left(\boldsymbol{x}\right),
\end{equation}
where $\boldsymbol{y} = \left[y_1, \cdots, y_K\right]^\mathrm{T}$ is a vector. The normalized bandwidth is updated by a readout function,
\begin{equation}
\boldsymbol{\tilde{W}} 
=f_\mathrm{W}(\boldsymbol{y})
= \boldsymbol{y} \tilde{W}_{\mathrm{B,S}}  + \tilde{\boldsymbol{W}}_{\min}.
\label{eq_vertex_feature}
\end{equation}

The parameters of the GNN is updated by stochastic gradient descent, i.e.,
\begin{equation}
\begin{split}
{\theta}^{t+1} ={\theta}^t - \beta \nabla_{\theta}L^{t}(\theta) 
,
\end{split}
\end{equation}
where $\beta$ is the learning rate, $ L^{t}(\theta)$ is the loss function in the $t$-th iteration. Parameter $\theta \in \{\theta_\mathrm{USL}, \theta_\mathrm{SL} \}$ is the trainable parameter of the neural network, where $\theta_\mathrm{USL}$ and $\theta_\mathrm{USL}$ represents the trainable parameters of the unsupervised and supervised learning algorithm, respectively.

\paragraph{Loss Function of Supervised Learning}
We use mean square error function, $f_\mathrm{MSE}(\cdot)$, as the loss function of GNN-SL, i.e.,
\begin{equation}
L(\theta_\mathrm{SL}) = f_\mathrm{MSE} \left( \tilde{\boldsymbol{W}}^\mathrm{*} -   \tilde{\boldsymbol{W}}\left(\theta_\mathrm{SL}\right)\right), 
\end{equation}
where $\tilde{\boldsymbol{W}}^*=[\tilde{W}_1^*, \tilde{W}_2^*,\cdots, \tilde{W}_K^*]$ is the optimal bandwidth allocation policy for the scheduled users, $\theta_\mathrm{SL}$ represents the parameters of the GNN-SL,

\paragraph{Loss Function of Unsupervised Learning}
The loss function is designed as
\begin{equation}
\begin{split}
L(\theta_\mathrm{USL}) 
&=\frac{1}{N_\mathrm{bat}}  \sum_{n=1}^{N_\mathrm{bat}} 
\sum\limits_{k=1}^{K} R_{n,k}^\mathrm{S} \left(W_{n,k}\left(\theta_\mathrm{USL}\right)\right) 
,
\end{split}
\end{equation}
where $\theta_\mathrm{USL}$ represents the trainable parameters in our designed GNN-USL, which is the $\theta$ in the $f_\mathrm{FNN}(\theta)$, $N_\mathrm{bat}$ is the batch-size. $W_{n,k}(\theta_\mathrm{USL})$ is the allocated bandwidth for the $k$-th scheduled user in the $n$-th batch, and is  determined by the GNN-USL. 
The secrecy rate of the $k$-th scheduled user is only related to the channels, and is given by
\begin{equation}
\begin{split}
R_k^\mathrm{S} (W_k(\theta_\mathrm{USL})) 
&={R}_k^\mathrm{S}\left(W_k(\boldsymbol{h}_k)\right)\\
&=R_k^\mathrm{B} (W_k(d_k^\mathrm{B}, g_k^\mathrm{B})) - R_k^\mathrm{E} \left(W_k(d_k^\mathrm{E} , g_k^\mathrm{E})\right).
\label{eq_positive_Rs}
\end{split}
\end{equation}

\subsection{Computational Complexity}
In this section, we analyse the computational complexity of our proposed GNN-USL approach with three benchmark schemes, namely GNN with supervised learning (GNN-SL), iterative search (IvS), and best channel (BeC). 
%
The following complexities are evaluated for processing one sample.

\paragraph{GNN-USL}
We consider the designed FNN has $L_\mathrm{FNN}$ layers, and the number of multiplications required to compute the output of the $\ell$-th layer is $m_\mathrm{FNN}^{\ell} \cdot m_\mathrm{FNN}^{\ell+1}$, where $m_\mathrm{FNN}^{\ell}$ is the number of neurons in the $\ell$-th layer. Since each user needs to use the FNN in each iteration, the number of multiplications for computing the FNN's output is $M_\mathrm{FNN} =  \sum\nolimits_{\ell=1}^{L_\mathrm{FNN}} m_\mathrm{FNN}^{\ell} \cdot m_\mathrm{FNN}^{\ell+1}$~\cite{SCY_transfer}.

In each iteration of the GNN, the multiplications required by $f_\mathrm{SOFTMAX}(\boldsymbol{x})$ and $f_\mathrm{W}(\boldsymbol{y})$ for individual users are both $1$. Since the rest operations, such as activation functions are much smaller, we can evaluate the complexity of training the GNN-USL bandwidth allocation as
\begin{equation}
O_\mathrm{GNN-USL}=O( K \cdot (M_\mathrm{FNN}+2+\Omega)).
\label{eq_complexity_USL}
\end{equation}

\paragraph{GNN-SL}
The complexity of the GNN-SL algorithm is the same as GNN-USL, and is given by
\begin{equation}
O_\mathrm{GNN-SL}
=O_\mathrm{GNN-USL} 
.
\label{eq_complexity_SL}
\end{equation}

However, we note that the GNN-SL algorithm also requires additional complexity for collecting the labels used in the model training which is not required for GNN-USL.

\paragraph{Iterative Search (IvS)}

In IvS, the number of iterations never exceed $\frac{W_\mathrm{B,S}}{\Delta W}$. Within each iteration, the IvS algorithm needs to compute $\Delta R_k^\mathrm{S}$ for each scheduled user with complexity $2\Omega$. Thus, the complexity using IvS to process a sample does not exceed 
\begin{equation}
O_\mathrm{IvS}=O\left(K \cdot\frac{W_\mathrm{B,S}}{\Delta W} \cdot  3 \cdot \Omega\right).
\label{eq_complexity_IvS}
\end{equation}

\paragraph{Best Channel (BeC)}
The BeC approach allocates all the surplus bandwidth, $W_\mathrm{B,S}=W_{\mathrm{B},\max} - \sum_{k=1}^{K} W_{k,\min}$, to the user with the best channel. Therefore, the complexity of BeC can be expressed as
\begin{equation}
O_\mathrm{BeC}=O( \Omega),
\label{eq_complexity_BeC}
\end{equation}
where $\Omega$ represents the multiplications required to compute each user's secrecy rate.

\textit{Comparison of Different Algorithms:}
%
We can observe from eqs.~\eqref{eq_complexity_USL},~\eqref{eq_complexity_SL}, and~\eqref{eq_complexity_IvS} that the computational complexity of GNN-USL, GNN-SL, and IvS all increase linearly with the number of scheduled users $K$, whilst in eq.~\eqref{eq_complexity_BeC} BeC is independent of $K$ since it always allocates the remaining bandwidth to the best user.
%
%
%
Furthermore, we note that the complexity of IvS also increases with smaller $\Delta W$ since it needs to allocate the $\Delta W$ for every iteration. In comparison, after the training, the GNN-USL and GNN-SL algorithms are both simply executed by a forward propagation operation based on a fixed size of the FNN.This means that the complexity of the learning algorithms can be designed to be much lower than the IvS algorithm especially when $\Delta W$ is very small.


\section{Numerical Evaluation}
\begin{table}[t] 
\renewcommand\arraystretch{1.1} 
\caption{Key Simulation Parameters} 
\centering 
\begin{tabular}{l l}
\toprule 
\toprule 
\textbf{Simulation parameters}&\textbf{Values}\\
\toprule
Transmit power of each user $P_u$                            &23 dBm~\cite{TWC_max_sum_rate}\\
\hline
Total bandwidth of the BS $W_{\mathrm{B},\max}$                        &10 MHz~\cite{TWC_max_sum_rate}\\
\hline
Number of total users $U$                                                  &10\\
\hline
Single-sided noise spectral density                        &-174 dBm/Hz\\
\hline
Threshold of secrecy rate $R_{\min}^\mathrm{S}$                               &0.8 Mbps\\
\hline
Size of resource block $\Delta W$                                      & 0.1 MHz\\
\hline
Learning rate $\beta$                                                         &$1\times 10^{-3}$\\ 
\hline
Batch size $N_\mathrm{bat}$                                                         &64\\ 
\bottomrule 
\bottomrule 
\end{tabular} 
\label{tab: Simulation Parameters} 
\end{table}

We consider the BS, the legitimate users, and the eavesdropper to be located within a square area, where the coordinates of the BS are $(0,0)$, and the coordinates of the legitimate users and the eavesdropper are generated randomly between (-100,-100) and (100,100). The small-scale fading coefficients of the legitimate users and the eavesdropper follow an independent and identically distributed (i.i.d.) Rayleigh distribution such that the channel gains $g_u^\mathrm{B}$ and $g_u^\mathrm{E}$, are exponentially distributed random variables with parameter $1$. We assume the eavesdropper frequently changes its location with dynamic coordinates and small-scale channel gains of the eavesdropper generated in each sample. Due to the dynamics of the wireless channels, the number of scheduled users are different in different training epoch and testing samples.
%
%
The designed FNN has three hidden layers, and the number of neurons in each layer, including the input layer, hidden layers, and output layer, are 2, 16, 8, and 1, respectively. For the SL algorithms, we generate $10^5$ samples of $\boldsymbol{h}_u, \forall u\in \mathcal{U}$. The IvS algorithm is executed to collect the training labels, and utilize the mean square error function as the loss function. Unless otherwise mentioned, the simulation parameters are summarized on Table.~\ref{tab: Simulation Parameters}.

\begin{figure}[t]
\centering
{\includegraphics[width=9cm, height=5cm]{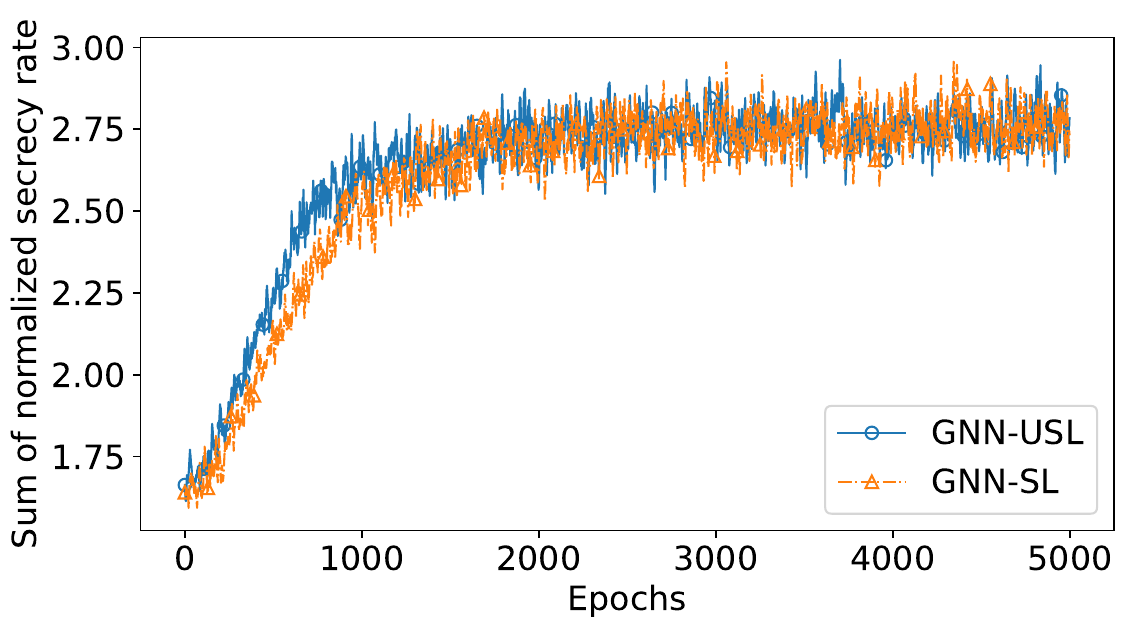}}
\caption{Training results of normalized average sum secrecy rate.}
\label{fig_trainig_loss}
\end{figure}
Fig.~\ref{fig_trainig_loss} shows the normalized average sum secrecy rate achieved by training with USL and SL algorithms. We can observe from this figure that both USL and SL are converged, and USL outperforms SL in terms of convergence speed. This is because the loss function of the SL algorithm is non-linear, which leads to a system error when comparing with the labels~\cite{SCY_tutorial_urllc}. In contrast, the loss function of unsupervised learning is the negative objective function without any system errors, thus does not have non-linearity problem.

\begin{figure}[t]
\centering 
%
%
{\includegraphics[width=9cm, height=5cm]{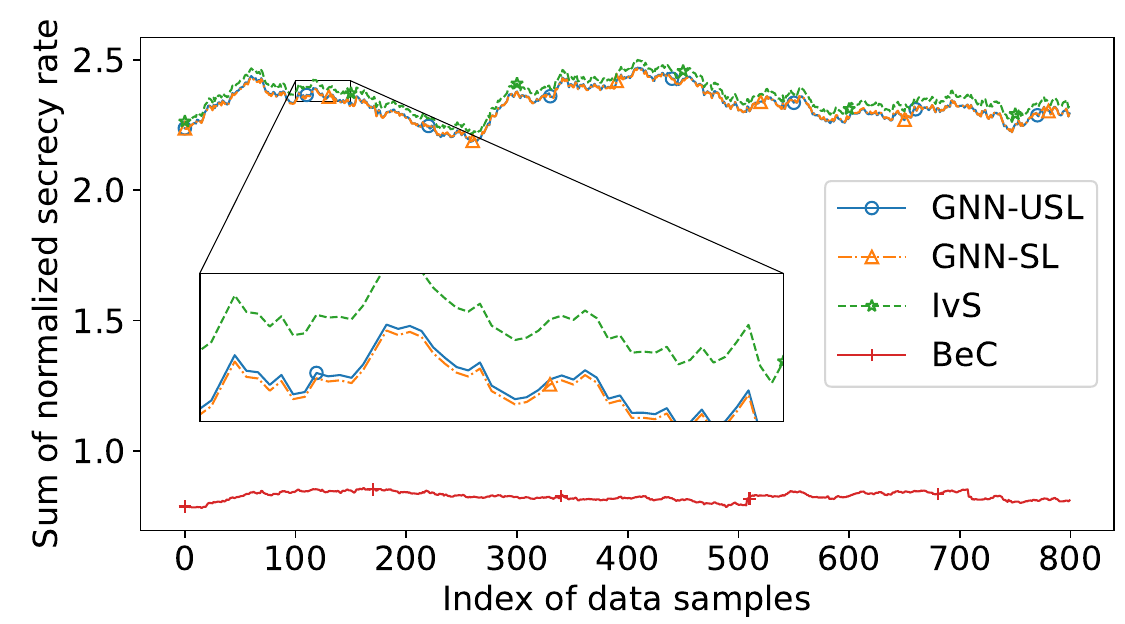}}
\caption{Normalized moving average sum secrecy rate using different bandwidth allocation algorithms.}
\label{fig_SumRs_ALL}
\end{figure}
In Fig.~\ref{fig_SumRs_ALL}, we generate another $10^3$ testing samples of $\boldsymbol{h}_u, \forall u \in \mathcal{U}$, to calculate the sum secrecy rate. The GNN-USL and GNN-SL algorithms use the well-trained neural networks achieved in Fig.~\ref{fig_trainig_loss}. We can observe from this figure that BeC has the worst average sum secrecy rate. In contrast, IvS has the best performance at the cost of high computational complexity with $\Delta W = 0.1$MHz. Within this testing data set, the average sum secrecy rate achieved by GNN-USL is approximately $98.7\%$ of IvS. 


\begin{figure}[t]
{\includegraphics[width=9cm, height=5cm]{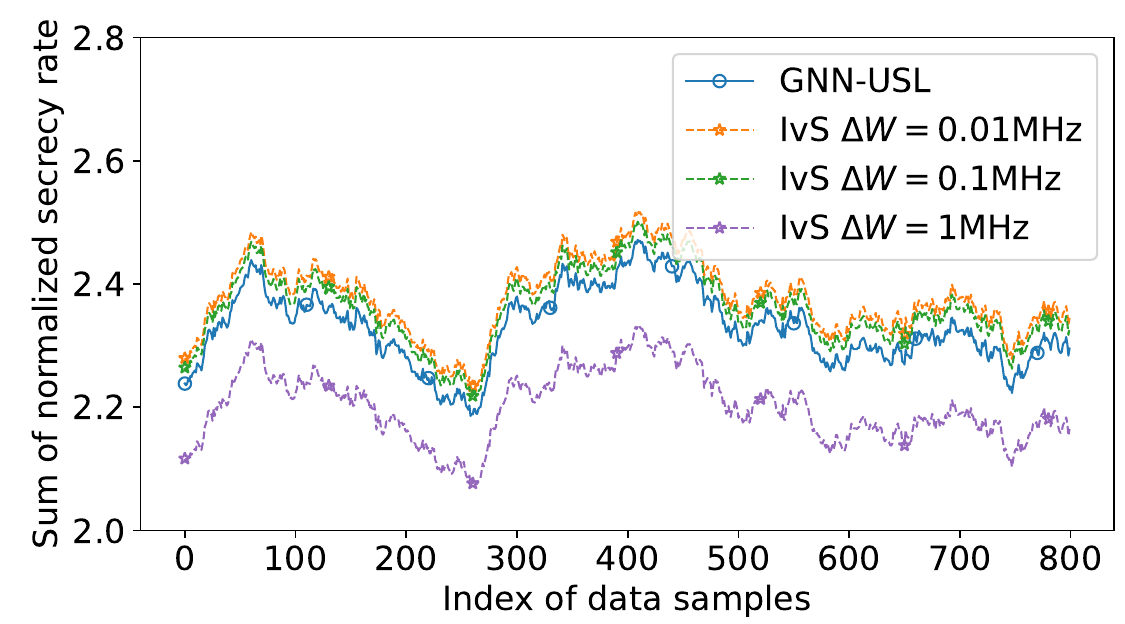}} 
\caption{Sum secrecy rate achieved by bandwidth algorithms with different complexities.}
\label{fig_complexity}
\end{figure}
Fig.~\ref{fig_complexity} shows the sum secrecy rate achieved using GNN-USL and IvS with different values of $\Delta W$. This figure shows that the sum secrecy rate achieved by IvS is highly related to $\Delta W$ as we discussed before. According to eq.~\eqref{eq_complexity_IvS}, we can reduce the complexity of IvS by approximately 10 times when $\Delta W$ is increased from 0.1MHz to 1MHz. However, the sum secrecy rate also suffers a drastically decrease. We note that when $\Delta W = 1$MHz, the complexity of GNN-USL is significantly less than IvS if $\Omega$ requires approximately 10 multiplications, which is a reasonable value. Even though $\Omega$ requires less than 10 multiplications for implementation, the complexity is comparable. {We can also observe from this figure that the sum secrecy rate only achieves limited incremental when we decrease $\Delta W$ from $0.1$MHz to $0.01$MHz, but increases the computational complexity for 10 times. Thus, $\Delta W=0.01$MHz is not a preferred value in our model for bandwidth allocation.} This figure demonstrates that GNN-USL can significantly reduce the complexity at the loss of a tolerable sacrifice on the sum secrecy rate compared with IvS.

\begin{figure}[t]
%
{\includegraphics[width=9cm, height=5cm]{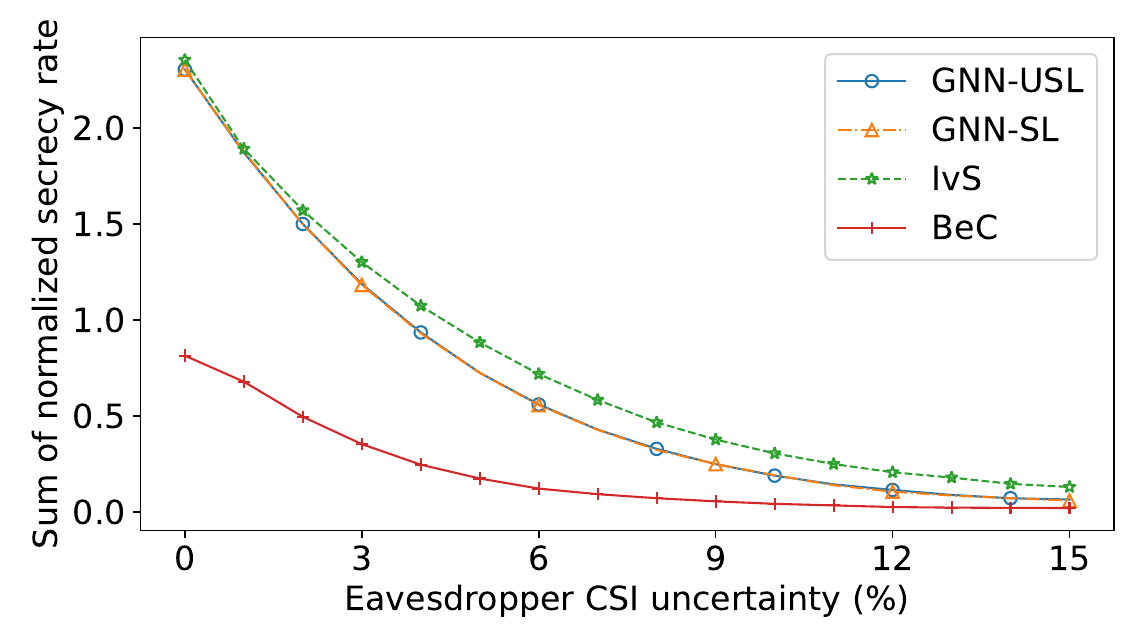}} 
\caption{Performance loss of sum secrecy rate when considering the eavesdropper's CSI uncertainty.}
\label{fig_Eve_uncertainty_Rs_trend}
\end{figure}
Fig.~\ref{fig_Eve_uncertainty_Rs_trend} depicts the average sum secrecy rate when introducing uncertainty into the eavesdropper's CSI. Since all four algorithms use the same user scheduling policy, we only consider the uncertainty of the eavesdropper's CSI in bandwidth allocation algorithms. The CSI uncertainty is defined as additive noise measured by a percent of $d_k^\mathrm{E}$ and $g_k^\mathrm{E}$. We can observe from this figure that all four bandwidth allocation algorithms decrease with increasing CSI uncertainty. Similar to Fig.~\ref{fig_SumRs_ALL}, we see that the GNN-USL approach can achieve a comparable sum secrecy rate to IvS and outperforms BeC across the range of different eavesdropper CSI uncertainty. Interestingly, GNN-USL can still achieve a significant improvement in sum secrecy rate compared to BeC even with moderate levels of CSI uncertainty. As expected, the performance of GNN-USL is close to BeC when the eavesdropper CSI uncertainty is high, {and if the CSI uncertainty is as large as $15\%$, the sum secrecy rates of all four algorithms decrease to approximately 0}. 


\section{Conclusion}
This paper proposed a graph neural network (GNN) based bandwidth allocation optimization problem to improve physical-layer security in a wireless network, where multiple legitimate users transmit confidential information to a BS in the presence of an eavesdropper with uncertain locations. To eliminate eavesdropping attacks, an instantaneous constraint on the secrecy rate is satisfied by the proposed user scheduling algorithm. A GNN supporting a dynamic number of neurons was designed to represent the dynamic wireless communication network, and {GNN-based SL and USL algorithms} were utilized to maximize the sum secrecy rate of all the users. Simulation results highlight the impact of uncertainties in the eavesdropper's channel state information (CSI) on the sum secrecy rate. Our proposed GNN-SL and GNN-USL algorithms can significantly reduce the computational complexity compared with existing benchmarks whilst achieving a comparable sum secrecy rate to the iterative search approach. {In future works, we plan to extend our current GNN to be a more robust model targeting the issues of wireless channel mismatches in the training and testing stage, and also extend the current basic GNN model adapting to scalable wireless networks.} 

\appendices 


\section{Proof of Concavity~\label{appendix_concave}}
Taking~\eqref{eq_user_data_rate} and~\eqref{eq_eavesdropper_rate} into~\eqref{eq_positive_Rs}, then we can calculate the partial derivative of the $k$-th scheduled user's secrecy rate as
\small
\begin{equation}
\begin{split}
\frac{\partial \hat{R}_k^\mathrm{S}(W_k)}{\partial {W_k}}
=&\frac{\partial  (R_k^\mathrm{B} (W_k)- R_k^\mathrm{E}(W_k) ) }{\partial W_{k}}\\
=&\frac{\ln\left(\frac{W_k + \xi_k^\mathrm{B}}{W_k+\xi_k^\mathrm{E}}\right) }{\ln(2)} 
+\frac{ (\xi_k^\mathrm{E} - \xi_k^\mathrm{B} ) W_k}{\ln(2)(W_k+\xi_k^\mathrm{B}  )(W_k+\xi_k^\mathrm{E})}
,
\end{split}
\end{equation}
\normalsize
where $\xi_k^\mathrm{B} = \frac{P_k (d_k^\mathrm{B})^{-\alpha}g_k^\mathrm{B}}{N0}$ and $\xi_k^\mathrm{E} =\frac {P_k (d_k^\mathrm{E})^{-\alpha}g_k^\mathrm{E}}{N0}$. Since the secrecy rate of a scheduled user increases with the increasing of the allocated bandwidth, we have $\frac{\partial \hat{R}_k^\mathrm{S}(W_k)}{\partial {W_k}}>0$. 

The second derivative of the $k$-th scheduled user's secrecy rate as
\small
\begin{equation}
\begin{split}
\frac{\partial^{2} \hat{R}_k^\mathrm{S}(W_k)}{\partial {W_k}^2}
=&\frac{\partial  }{\partial W_k} \left(\frac{\partial \hat{R}_k^\mathrm{S}(W_k)}{\partial {W_k}}\right)\\
=&\dfrac{(\xi_k^\mathrm{E} - \xi_k)\left((\xi_k^\mathrm{E} + \xi_k)W_k + 2\xi_k^\mathrm{E} \xi_k\right)}
{\ln(2) (W_k + \xi_k)^2 (W_k + \xi_k^\mathrm{E})^2}
.
\end{split}
\end{equation}
\normalsize
For the scheduled legitimate users, since we have $\xi_k  > \xi_k^\mathrm{E}$, thus $\frac{\partial^{2} \hat{R}_k^\mathrm{S}(W_k)}{\partial {W_k}^2} < 0$. In addition, when $j \neq k$, we have $\frac{\partial \hat{R}_k^\mathrm{S} \partial \hat{R}_j^\mathrm{S}}{\partial {W_k} W_j}=0$. Thus, the Hessian matrix of function $\sum_{k =1}^{K}  \hat{R}_k^\mathrm{S} (W_k(g_k^\mathrm{B}, g_k^\mathrm{E})$ is semi-negative definite, and problem~\eqref{eq_positive_Rs_basic} is concave. This completes the proof. {\hfill $\square$\par}

\appendices 


\begin{thebibliography}{00}
%
\bibitem{SCY_tutorial_urllc} C. She, C. Sun, Z. Gu, Y. Li, C. Yang, H. V. Poor, and B. Vucetic, ``A tutorial on ultrareliable and low-latency communications in 6G: Integrating domain knowledge into deep learning,'' \emph{Proc. IEEE}, vol.~109, no.~3, pp.~204--246, Mar.~2021. 
%
\bibitem{SCY_digital_twin} R. Dong, C. She, W. Hardjawana, Y. Li, and B. Vucetic, ``Deep learning for hybrid 5G services in mobile edge computing systems: Learn from a digital twin,'' \emph{IEEE Trans. Wireless Commun.}, vol.~18, no.~10, pp.~4692--4707, Oct.~2019.
%
%
%
\bibitem{outperforms_random_SVM_etc} R. Yao, Y. Zhang, S. Wang, N. Qi, N. I. Miridakis, and T. A. Tsiftsis, ``Deep neural network assisted approach for antenna selection in untrusted relay networks,'' \emph{IEEE Wireless Commun. Lett.}, vol.~8, no.~6, pp.~1644--1647, Dec.~2019.
%
%
%
\bibitem{LYH_GNN_interference} Y. Liu, C. She, Y. Zhong, W. Hardjawana, F.-C. Zheng, and B. Vucetic, ``Interference-limited ultra-reliable and low-latency communications: Graph neural networks or stochastic geometry?'', \emph{arXiv preprint arXiv:2207.06918} [Online]. Available: https://arxiv.org/abs/2207.06918
%
\bibitem{industry_Secure_ML} H. Ren, C. Pan, Y. Deng, M. Elkashlan, and A. Nallanathan, ``Resource allocation for secure URLLC in mission-critical IoT scenarios,'' \emph{IEEE Trans. Commun.}, vol. 68, no. 9, pp. 5793-5807, Sept. 2020.
%
%
\bibitem{Ring_shape_location_Eve} D. S. Karas, A. A. Boulogeorgos, and G. K. Karagiannidis, ``Physical layer security with uncertainty on the location of the eavesdropper,'' \emph{IEEE Wireless Commun. Lett.}, vol.~5, no.~5, pp.~540--543, Oct.~2016.
%
%
\bibitem{TWC_SR_Poor} W. Yu, A. Chorti, L. Musavian, H. Vincent Poor, and Q. Ni, ``Effective secrecy rate for a downlink NOMA network,'' \emph{IEEE Trans. Wireless Commun.}, vol.~18, no.~12, pp.~5673--5690, Dec.~2019.
%
\bibitem{thz_SR_GC} W. Gao, C. Han, and Z. Chen, ``Receiver artificial noise aided terahertz secure communications with eavesdropper in close proximity,'' in \emph{Proc. Global Commun. Conf. (GLOBECOM)}, Taipei, Taiwan, 2020, pp.~1--6.
%
\bibitem{tvt_RS_non_robust} M. Zhang, K. Cumanan, J. Thiyagalingam, Y. Tang, W. Wang, Z. Ding, and O. A. Dobre, ``Exploiting deep learning for secure transmission in an underlay cognitive radio network,'' \emph{IEEE Trans. Veh. Technol.}, vol.~70, no.~1, pp.~726--741, Jan.~2021.
%
\bibitem{INFOCOM_SR_DRL} H. Sharma, I. Budhiraja, N. Kumar, and R. K. Tekchandani, ``Secrecy rate maximization for THz-enabled femto edge users using deep reinforcement learning in 6G,'' in \emph{IEEE Conf. on Comput. Commun. Workshops (INFOCOM WKSHPS)}, New York, NY, USA, 2022, pp.~1--6.
%
%
%
%
%
%
%
%
%
\bibitem{TWC_max_sum_rate} C. Guo, L. Liang, and G. Y. Li, ``Resource allocation for vehicular communications with low latency and high reliability,'' \emph{IEEE Trans. Wireless Commun.}, vol.~18, no.~8, pp.~3887--3902, Aug.~2019.
%
\bibitem{MPNN_chemistry} J. Gilmer, S. S. Schoenholz, P. F. Riley, O. Vinyals, and G. E. Dahl, ``Neural message passing for quantum chemistry,'' in \emph{Proc. Int. Conf. Mach. Learn. (ICML)}, Sydney, Australia, pp.~1263--1272, Apr.~2017.
%
%
\bibitem{MPGNN_HK_JSAC} Y. Shen, Y. Shi, J. Zhang, and K. B. Letaief, ``Graph neural networks for scalable radio resource management: Architecture design and theoretical analysis,'' \emph{IEEE J. Sel. Areas Commun.	}, vol.~39, no.~1, pp.~101--115, Jan.~2021.
%
\bibitem{SCY_transfer} R. Dong, C. She, W. Hardjawana, Y. Li, and B. Vucetic, ``Deep learning for radio resource allocation with diverse quality-of-service requirements in 5G,'' \emph{IEEE Trans. Wireless Commun.}, vol.~20, no.~4, pp.~2309--2324, Apr.~2021.



%
%
%
%
%
%
%
%
\end{thebibliography}
\end{document}